\documentclass[aip,psfig,epsfig]{article}
\newcommand{\tu}{\widetilde}

\newcommand{\pa}{\partial}

\begin{document}

%\draft
\title{Quantum horizon and thermodynamics of black hole   }
\author{{Mu-Lin Yan\thanks{E-mail address: mlyan@staff.ustc.edu.cn},
 Hua Bai\thanks{E-mail address: huabai@mail.ustc.edu.cn}}\\
{\small Interdisciplinary Center for Theoretical Study}\\
{\small University of Science and Technology of China}\\
{\small Hefei, Anhui 230026, P.R. China}}
%\date{\today}
\maketitle

\begin{abstract}
%\begin{center}
%{\bf Abstract}
%\end{center}
%\vspace{-0.1in}
 A semi-classical reasoning leads to the
non-commutativity of space and time coordinates near the horizon
of static non-extreme black hole, and renders the classical
horizon spreading to {\it Quantum Horizon} . In terms of the
background metric of the black hole with the {\it Quantum Horizon}
, a quantum field theory in curved space without ultraviolet
divergency near the horizon is formulated. In this formulism, the
black hole thermodynamics is reproduced correctly without both
ambiguity and additional hypothesis in the deriving the hole's
Hawking radiations and entropies, and a new interesting prediction
on the number of radiative field modes  $N$  is provided.
Specifically, the main results are follows: 1, Hawking radiations
rightly emerge as an effect of quantum tunneling through the
quantum horizon, and hence the ambiguities due to going across the
singularity on the classical horizon were got rid of; 2, 't
Hooft's brick wall thickness hypothesis and the boundary condition
imposed for the field considered in his brick wall model were got
rid of also, and related physics has been interpreted; 3, The
present theory is parameter free. So, the theory has power to
predict the multiplicity $N$ of radiative field modes according to
the requirement of normalization of Hawking-Bekenstein entropy. It
has been found that $N\simeq 162$, which is just in good agreement
with one in the Minimal Super-symmetric Standard Model. The
studies in this paper represent an attempt to reveal some physics
near the horizon at Planck scale. This paper serves a brief review
on the author's works on this subject.

\end{abstract}

\maketitle
\section{INTRODUCTION}

  It is well known that the general relativity(GR) and the quantum theory are
two main pillars of modern physics. That how to  unite these two
theories is full of challenge. Actually, so far no one can yet
 formulated a consistent and complete quantum theory of
gravity. We all know that gravitational interactions govern the
phenomena at large scales, and quantum mechanics(QM) and quantum
field theory(QFT) dominates the physics at small scales, but once
the distance is small as Planck length, $l_p=(\hbar G/c^3)^{1\over
2}\approx 10^{-35} m$, we have to consider the quantum effect of
gravity.

Two of the most important physical effects related to the quantum
gravity are the Hawking-radiation and the black hole entropy. In
classical GR, black hole's properties can be precisely calculated,
and the holes may be thought of astronomical objects with masses
about several times of our Sun. In this classical case, event
horizon emerges, and anything (including the light) can not escape
from it to arrive at a particular observer who is outside the
horizon. But a surprise happened: When S. Hawking studied the
Bekenstein-relation of black hole, he found that black hole emits
radiations and the radiation spectra is just the black body's.
This fact indicates that the black hole has a temperature whose
expression is\cite{Hawking74}
\begin{equation}\label{tm}
T_{BH}={\kappa \over 2\pi},
\end{equation}
where $\kappa$ is surface gravity on the horizon. Thus the
Bekenstein-relation becomes the "real" thermodynamical
relationship of the black hole. Namely the black hole has a
thermodynamical entropy as follows
\begin{equation}\label{BHentropy}
S_{BH}={1\over 4} {A\over l_p^2},
\end{equation}
where $S_{BH}$ is the Bekenstein-Hawking entropy and $A$ is the
area of the event horizon.

The Quantum Mechanics(QM) and the Quantum Field Theory (QFT) in
the curved space-time with (classical) event horizon provides a
framework to understand the statistic mechanics origin of the
thermodynamics of non-extreme black hole, and serves as a powerful
tool to calculate its entropy and the Hawking radiation
temperature\cite{Hawking74}\cite{thooft04}\cite{wald}\cite{Ruffini}
\cite{thooft85}\cite{thooft96}. In this framework, the gravity
fields are  background fields. The quantum fields move on this
classical background which will be not affected  by the quantum
fields. It is essential that the black hole background has
coordinate singularity at the horizon. This singularity will
seriously plague our understanding of the black hole's
thermodynamics:Firstly, in the entropy calculations, an
ultraviolet cutoff (or brick wall) has to be put in by
hand\cite{thooft85} (even though it could be done in a proper way
\cite{thooft04}). Otherwise, the entropy will be divergent;
Secondly, in the Hawking radiation derivations, one should also
artificially use an analytic continuing trick to go over the
singularity associating with that
horizon\cite{Hawking74}\cite{Ruffini}\cite{Wilczek}. This disease
spoils our understanding of Hawking radiation as quantum
tunnelling effects, and some ambiguities will be left due to it.
Thus we face an unsatisfactory situation that if once set up a
brick wall for getting the entropy rightly, we will have no way to
do analytic continuing and hence no way for deriving the Hawking
temperature in this formalism. On the other hand, if once withdrew
the wall (or the ultraviolet cutoff), we will have no way to
handle the entropy even though the Hawking temperature deriving
become doable. Consequently, a self-consistent QFT (or QM) model
in curved space to calculate both the black hole's entropy and the
Hawking temperature simultaneously and rightly is still in
absence.

 Recently, we made some remarks\cite{by03}\cite{by04} on 't Hooft's
brick wall model(BWM) which is one of QFTs in curved space with
(classical) horizon, and provided a heuristic discussion for how
to over the obstacle mentioned in above. The point of our idea is
to consider a semi-classical effect on the horizon , which
naturally led to the space-time non-commutativity on the horizon,
and dramatically and successfully led to over the difficulties due
to the singularity at the classical horizon. We will call such
kind of event horizon on which the space-time is non-commutative
as {\it Quantum Horizon} in this paper.  As a consequence in
ref.\cite{by03}\cite{by04}, we simultaneously got the
Schwarzschild black hole's Hawking temperature and entropy in a
probing QFT-model with a quantum horizon. In this treatment, the
ordinary classical horizon spreads, and hence the troubles  due to
the singularity on the horizon mentioned in above disappear. This
seems to be a strong sign to indicate that the quantum effects on
the horizon are important for the the thermodynamics of black
hole. In other wards, the quantum horizon conception may play a
essential role for solving the puzzle brought by the classical
horizon singularity.

Following refs.\cite{thooft04}\cite{thooft96} to treat macro-black
hole as quantum states instead of a classical object, and
according to quantum mechanics principle, then we should conclude
its energy and its corresponding conjugate time $t$ can not be
simultaneously measured exactly (Heisenberg uncertainty
principle). Namely, when treating $E$ and $t$ as operators, we
have
\begin{equation}\label{Heisenb}
[t, E]=i.
\end{equation}
In other hand, it has been generally believed that the states are
situated on the horizon, and outside it the densities of the
states become tiny fast. Therefore, the energy of states $E$
should approximatively be the hole's gravitational energies
measured at the location of the horizon. It has long been aware
the gravitational energies are quasilocal in the general
relativity. In this letter, we follow Brown-York's quasilocal
energy(QLE) defination\cite{Brown93}, and the energy of the black
hole $E_{QLE}$ is coordinate-dependent. Then, the states energy
$E$ to the static black hole is as follows(see the  Sectiion III)
\begin{equation}\label{1}
E=E_{QLE}(r=r_H)=r_H/G,
\end{equation}
where $r_H$ is the radio of black hole (or the radio of event
horizon), and it is a function of the parameters of the
hole.Substituting  Eq.(\ref{1}) into Eq.(\ref{Heisenb}), we must
conclude that the uncertainty of $E$ imply that of $r_H$, and then
we have
\begin{equation}\label{NC}
[t,r]|_{r=r_H}=il_p^2,
\end{equation}
where $l_p=\sqrt{G}$ is the Planck length. This equation implies
that the radial coordinate is noncommutative with the time at the
horizon. The corresponding uncertainty relation for them is
$(\Delta t)(\Delta r)|_{r \sim r_H} \sim l_p^2$. In other hands,
due to quantum measurement effects, $r_H$ spreads into a range of
$\{r_H-\Delta r, r_H+\Delta r\}$. And correspondingly, the
previous classical horizon spreads into the quantum horizon. Thus
Eq.(\ref{NC}) is extended as follows
\begin{equation}\label{NC1}
[t,r]|_{r\in \{r_H-\Delta, r_H+\Delta\}}=il_p^2,
\end{equation}
According to our aforementioned discussion, we construct a model
with noncommutative $\phi$-field in the range of $\{r_H -\Delta,
r_H + \Delta\}$ and with commutative $\phi$-fields in the ranges
of $\{0, r_H -\Delta \}$ and  $\{r_H +\Delta,L\}$
\begin{equation}\label{t}
[t,r]|_{r\in\{0, r_H -\Delta \} \;\; or \;\; \{r_H +\Delta,L\}}=0
\end{equation}
where $L$ represents an infrared cutoff in the model. It is
essential that $\Delta$ should be an intrinsic quantity of the
model, which characterizes the boundary between the noncommutative
space-time range and the commutative space-time ranges, and should
be determined by dynamics of the model. Surprisingly, this
expectation can be realized and a ultraviolet-divergence-free QFT
with horizon but without the "brick wall" can be constructed by
the following consideration: 1) Starting with a simplest
noncommutative $\phi$-field action within metric of black hole,
the equation of motion of $\phi$ can be derived exactly; 2) This
equation of motion in noncommutative field theory should be, of
course, quite different from the ordinary Klein-Gordon equation of
$\phi$-field within black hole metric. This fact implies
$\phi$-fields should be moved in a curve space with a new
effective metric $\widetilde{g}^{\mu\nu}$;3) Remarkably, it will
be shown below that $\widetilde{g}^{tt}$ has two new singularities
besides the the original one at $r=r_H$, one is outside the
horizon and another is in inside. Denoting their locations as $r_H
\pm \Delta '$ respectively, we will find $\Delta '$ is dependent
on the energies of the noncommutative fields $\omega$,i.e.,
$\Delta '=\Delta '(\omega)$. This means that to the fields with
energy $\omega$, $\phi(\omega,r)$, its red-shifting on the
$(r_H\pm \Delta '(\omega))$ surface is infinite due to
$\widetilde{g}_{tt}(r=r_H\pm\Delta' (\omega))=0$, and then we have
\begin{equation}\label{bc}
\phi(\omega,r) |_{r=r_H + \Delta ' (\omega)}=0;
\end{equation}
4) We argue that the fact that the noncommutative fields vanish at
$r=r_H + \Delta '(\omega)$ means we can think the space-time
coordinates on the surface of $r=r_H + \Delta '(\omega)$ to be
commutative, i.e., $[t,r]|_{r=r_H + \Delta '(\omega)}=0$. and then
we will further have
\begin{equation}\label{t1}
[t,r]|_{r\geq r_H +\Delta '(\omega)}=0.
\end{equation}
Comparing eq.(\ref{t1}) with eq.(\ref{t}), we get
\begin{equation}\label{bc1}
\Delta ' (\omega)=\Delta .
\end{equation}
Because of the appearance the new infinite red-shifting surface,
an observer outside this surface can not detect the $\phi_\omega$
at $r<r_H+\Delta(\omega)$, so the state density divergency of the
$\phi$-field caused by the coordinate singularity at the original
horizon($r=r_H$) disappears. Thus, in this theory without
introducing  extra-violet cutoff (or brick wall) the entropy  will
keep to be finite. It interesting to pursue whither this entropy
 is proportional to the area of the black hole or not, or if the black hole
thermodynamics could be reproduced or not. If does, the quantum
horizon
 would be certainly correct conception, and our treatment
 should be a right way to regulate the divergent at the horizon,
 which may
  be more nature than the 't Hooft brick wall hypothesis.

 In the horizon
spread range, the wave number $k$ of the $\phi$-fields is
imaginary, and hence the quantum tunnelling occurred. The ratio
between the amplitude of outgoing wave function of $\phi$ outside
the hole and one inside it is
$\exp({2i\int^{r_H-\Delta}_{r_H+\Delta} k(r)dr})$, which describes
the the quantum tunnelling effects of $\phi$ passing through the
horizon spread range of the black hole. By means of the theory of
Dammur-Ruffini\cite{Ruffini} and Sannan\cite{Sannan}, the black
hole temperature $T_{BH}$ can be derived from this ratio , i.e.,
$\exp(|{2i\int^{r_H-\Delta}_{r_H+\Delta} k(r)dr}|)=\exp (\omega
/(2T_{BH}))$, and then it becomes possible to examine whither the
Howking radiation is  due to this well-define quantum tunnelling
effects or not. It should be remarkable to link the black hole
thermodynamics with the quantum effects at the horizon.

Finally, we discuss the question of the entropy normalization. The
thermodynamics requests the hole's entropy have to normalize to be
Bekenstein-Hawking expression(\ref{BHentropy}), i.e.,
$S=S_{BH}=A/(4l_p^2)$. This requirement will lead to determine the
$\phi-$field component number $N$ both because $S\propto N$ in
general and because, in particular,  there is no free parameter in
our model. Therefore distinguishing from other type  brick wall
models, our quantum horizon model can predict the $N$. We will
find $N\approx 162$. This result provides a important criterion to
judge what underline dynamics at the Planck scale should be.

The contents  of this paper are organized  as follows. In section
II we  briefly review some relevant works, and reveal the puzzles
related to the singularities at the (classical) horizons in the
QFT (or QM) on the curved space; In section III we discuss the
energies of
  quantum state of the black hole. By means of the Heisenberg uncertainly
  principle we will show
that the space-time near the event horizon is noncommutative ; In
 section IV,  a model based on the quantum
horizon will constructed, and  the thermodynamics of the
Schwarzschild black hole will be produced in this model; In
section V,  our studies on the Schwarzschild black hole extend to
the general static black holes . It will be shown that our model
is very successful for getting both the temperature and entropy of
the black hole without any free parameters. The model will also
successfully lead  to get over the difficulties due to the
singularity at the classical horizon discussed in the section II;
Finally, we summarize the results, and discuss the question of the
entropy normalization in our model.

\section{QFT  on the curved space-time }
In order to see the problems of the QFT(or QM) on the curved space
with horizon , we first briefly recall some relevant works in this
field , for instance,
 the works
of Damour-Ruffini\cite{Ruffini},Sannan\cite{Sannan} and t'
Hooft\cite{thooft04}\cite{thooft85}\cite{thooft96} . To the
Schwarzschild black hole, the metric is given by
\begin{equation}\label{Schwarzchild}
ds^2=-(1-{2M\over r})dt^2+(1-{2M\over
r})^{-1}dr^2+r^2(d\theta^2+\sin^2(\theta)d\varphi^2).
\end{equation}
Considering a scale field in this background, the action is as
follow
\begin{equation}\label{act}
I=-{1\over 2}\int d^4x \sqrt{-g}g^{\mu\nu}(\pa_\mu \phi \pa_\nu
\phi)
\end{equation}
Varying the action with $\phi$, i.e., $\delta I/ \delta \phi$, we
get the equation of motion as follow
\begin{equation}\label{KG}
\pa_\mu(g^{\mu \nu}\pa_\nu \phi)=0
\end{equation}
Using the WKB approximation and by setting $\phi=e^{-i\omega
t-i\int k(r)dr}Y_{lm}(\theta ,\varphi)$, we have the radial wave
number $k(r,l,\omega)$ from the corresponding equation of motion:
\begin{equation}\label{kk}
k^2(r,l,\omega)=(1-{r_H\over r})^{-1}\left[ (1-{r_H\over
r})^{-1}\omega^2-r^{-2}l(l+1)\right],
\end{equation}
where $r_H=2M$ is the ratio of the event horizon. Now, let us
derive the temperature of the black hole firstly. Using the S-wave
approximation corresponding to eq.(\ref{k}), we have
\begin{equation}\label{k}
k(r)=\pm {r\omega\over r-r_H}
\end{equation}
We should derive the outgoing wave function of $\phi$ in both
outside and inside of the black hole. The incoming wave function
of $\phi$ reads
\begin{equation}
\phi^{in}_\omega = \exp \left[ -i\omega t-i\int k(r)dr
\right]=\exp \left[ -i\omega (t+\int^r_c {r\over r-r_H}dr)\right]
.
\end{equation}
In the above expression, the integration $\int^r_c r/(r-r_H)
dr=r-c+r_H\ln ((r-r_H)/(c-r_H))$ is actually of the so called the
 tortoise coordinate $r^\ast=r+r_H\ln ((r-r_H)/r_H)$ used in
ref.\cite{Ruffini}, expect an unimportant constant. We set
$t+\int^r_c r/(r-r_H) dr=\nu $ which is like the usual advanced
Eddingtion-Finkelstein coordiantes. Then the incoming wave
function is $\phi^{in}=e^{-i\omega \nu}$, and the outgoing wave
function in the outside range $(r>r_H)$ is
\begin{eqnarray}
\phi^{out}_\omega(r>r_H) &=& A_\omega\exp \left[ -i\omega
t+i\int^r_c {\omega r\over r-r_H}dr \right]   \\ \nonumber
 &=& A_\omega\phi^{in}_\omega\exp
\left[ 2i\int^r_c {\omega r\over r-r_H}dr\right]
\end{eqnarray}
The outgoing wave function has a singularity at $r=r_H$ in the
integrand, if we want to get the outgoing wave function in the
inside range$(r<r_H)$, we have to do analytic continuing over this
singularity. Namely, the calculations of imaginary part of the
wave-number $k$ integral across the classical horizon are as
follows
\begin{equation}\label{im}
Im \left( 2\int dr k(r)\right) \equiv Im\left(2\int {\omega r\over
r-r_H}dr \right)=Im\left(2\omega \lim_{\epsilon\rightarrow
0}\int{rdr\over r-r_H\mp i\epsilon}\right)=\pm2\pi r_H \omega
\end{equation}
Note, the calculations here are somehow non-rigorous. Two remarks
on this point are follows: 1) The integration $\int k(r)dr$ is
actually divergent due to the singularity at $r=r_H$ in the
integrand, and hence $Im(2\int k(r)dr)$ is actually meaningless as
without a proper regularization; 2), The second step of
calculation (\ref{im}) just serves as such a regularization which
belongs to a artificial mathematical trick, and the final step is
due to the Dirac formula: $1/(x\pm i\epsilon)=\mathcal{P} / x \mp
i\pi\delta(x)$.

Now, the outgoing wave function inside the horizon is
\begin{equation}
\phi^{out}_\omega(r<r_H)=A_\omega e^{2\pi r_H \omega}
\phi^{in}_\omega\exp \left[ 2i\int {\omega r\over r-r_H}dr\right]
\end{equation}
where for simplicity, we take the positive sign(if taking the
negative sign, we can get the same result\cite{Birrell}). The
absolute value of ratio of the outgoing wave function's amplitude
outside the black hole to the one inside is as follows
\begin{equation}\label{ratio}
\left| {\phi^{out}_\omega(r>r_H)\over
\phi^{out}_\omega(r<r_H)}\right| = e^{-2\pi r_H \omega}
\end{equation}
This yields a relative scattering probability
\begin{equation}\label{ratio1}
p_\omega=e^{-4\pi r_H \omega}
\end{equation}
According to Feynman \cite{Feynnam}, the probability $p_\omega$ is
also the relative probability of creating a particle-antiparticle
pair just outside the horizon. The particles are outgoing from the
event horizon with positive energy $\omega$, while the
antiparticle is ingoing toward the singularity with energy
$(-\omega)$. This means that the $\phi$-particle mean number
$\langle\mathcal{N}_\omega\rangle$ in the radiation is as
follows\cite{Sannan}
\begin{equation}\label{ratio2}
\langle\mathcal{N}_\omega \rangle ={|\Gamma_\omega |^2\over
e^{4\pi r_H \omega}-1}\equiv {|\Gamma_\omega |^2\over e^{\omega /
T_{BH}}-1},
\end{equation}
where $|\Gamma_\omega |^2$ is the frequency-dependent transmission
coefficient for the outgoing wave to reach future infinity.
Consequently, we obtain the Hawking temperature
\begin{equation}\label{TBH}
T_{BH}={1\over 4\pi r_H}.
\end{equation}
To this temperature calculation, the regularization in (\ref{im})
is crucial, but {\it ad hoc}. Actually, some ambiguities are left.
For example, the regularization used in (\ref{im}) could be
replaced by the follows
\begin{eqnarray}\label{im1}
&&Im\left( 2\int dr k(r)\right) \equiv Im\left(2\int {\omega
r\over r-r_H}dr \right)\\ \nonumber &=&Im\left(2\omega
\lim_{\epsilon\rightarrow 0}{1\over m+n} \left[m \int{rdr\over
r-r_H- i\epsilon}+n \int{rdr\over r-r_H+ i\epsilon}\right]
\right)\\ \nonumber &=&\pm2 {m-n \over m+n}\pi r_H \omega
\end{eqnarray}
where $m$ and $n$ are arbitrary integers. In this case the
radiation temperature becomes
\begin{equation}\label{TBH1}
T_{BH}=|{m+n \over m-n}|{1\over 4\pi r_H}.
\end{equation}
which is obviously regularization-dependent, and is inconsistent
with the standard result (\ref{TBH}). This puzzle will be solved
in our quantum horizon model(see the  Section IV).

Now we turn to discuss the Brick Wall Model and the calculations
of the black hole's entropy. Since $k(r)$ has singularity at
$r=r_H$ and then the
   state density (which corresponds to the entropy) is divergent,
, t' Hooft suggest to put a "brick wall" near the
horizon\cite{thooft04} \cite{thooft85} \cite{thooft96}
\begin{equation}
\phi(r,\theta,\varphi,t)=0 \;\;\;\;\;\;{\rm
if}\;\;\;\;\;\;(r<r_H+h,\;\;\;\; {\rm or} \;\;\;\; r>L),
\end{equation}
where $L$ is large (infrared cutoff) and $h$ is the thickness of
the brick wall (corresponding to the extra-violet cutoff). In this
model, $h$ has been assumed to be as follows
\begin{equation}\label{h}
h={N'l_p^2\over 360\pi r_H}.
\end{equation}
We address this is a {\it prior} hypothesis of this model, and its
physical meanings are unclear.

 For simplicity, we only study the singlet
scale field in the above. It is straightforward to extend the
studies of actions of (\ref{act}) of $\phi$ to ones of $\phi^A$
with $A=1\cdots N'$ where $N'$ is the number of the fields. As
well known that all the fields couple with the gravity force which
emitting from the vacuum near the horizon, also, all of their
dynamical degrees freedom will contribute to entropy.
 For example, two-component
spin field contribution to the entropy is double of which of the
scale field. The number of states below energy $\omega$ is
\begin{equation}\label{g}
g(\omega)=N'\int dl(2l+1)\int^L_{r_H+h} dr\sqrt{k^2(r,l,\omega)},
\end{equation}
where $k^2$ is given by eq.(\ref{k}). The free energy $F$ then
reads
\begin{eqnarray}
\pi\beta F &=&\int dg(\omega) \ln \left( 1-e^{-\beta
\omega}\right)
\\  \nonumber &=& -\int^\infty_0 d\omega {\beta g(\omega)\over
e^{\beta \omega}-1},
\end{eqnarray}
where $\beta=1/T_{BH}$ is the inverse temperature. The dominant
contribution from the event horizon to $F$ is
\begin{equation}
F\approx -{2\pi^3\over 45 h}\left( {r_H\over \beta}\right)^4
\end{equation}
The contribution of the horizon to the entropy $S$ are
\begin{eqnarray}
 S= \beta^2{\pa F\over \pa \beta}={8\pi^3\over 45
h}r_H\left({r_H\over \beta}\right)^3N'.
\end{eqnarray}
Comparing the entropy $S$ with the Bekenstein-Hawking's result
Eq.(\ref{BHentropy}), and noting
  $\beta =4\pi r_H$, we can see that under the above brick wall
  thickness hypothesis eq.(\ref{h}) the black hole thermodynamics is
  reproduced.

\section{Quasilocal Energy}

According to the discussions  in the introduction, the energy of
the quantum states of the black hole is the QLE  on the horizon
\begin{equation}
E=E_{QLE}(r_H).
\end{equation}
 In this section, we study
the static spherical non-extreme black hole. Its metric can
generically  be expressed as follows
\begin{equation}\label{metric}
ds^2=-D(r)dt^2+D^{-1}(r)dr^2+r^2(d\theta^2+\sin^2\theta
d\varphi^2),
\end{equation}
where $D(r)=f(r)(r-r_H)$. The time Killing vector, $\pa_t$, is
everywhere. $r_H$ is the radius of the event horizon, and it
generically is the functions of the mass, charge, cosmical
constant of the black hole parameters.

 There are inherent
difficulties in defining energy in GR, essentially due to its
non-localizability. For QLE, we shall  adopt the definition given
by Brown and York\cite{Brown93}.
\begin{equation}\label{QLE}
E_{QLE}={1\over 8\pi G}\oint_B d^2 x \sqrt{\sigma} (K -K_0),
\end{equation}
where $B$ is the two dimension spherical surface, $\sigma$ is the
determinant of the 2-metric on $B$, $K$ is the trace of the
extrinsic curvature of $B$, and $K_0$ is a reference term that is
used to normalize the energy with respect to a reference
space-time, not necessarily flat. To the given metric
eq.(\ref{metric}), $K$ is given by
\begin{equation}
K^\theta_\theta=K^\varphi_\varphi=-{\sqrt{D(r)}\over r}.
\end{equation}
For space-times that are asymptotically flat in spacelike
directions, the QLE eq.(\ref{QLE}) with $B$ at spatial infinity
agrees with the ADM energy\cite{ADM}. Now, we chose the $K_0$ is
 as follows
\begin{equation}
K_0=-{1\over r}.
\end{equation}
The QLE of the black hole reads
\begin{equation}\label{39}
E(r)_{QLE}=r(1-\sqrt{D(r)})/G
\end{equation}
Specifically, for Reissner-Nordstr\"{o}m black hole,
\begin{equation}
D(r)={r- r_- \over r^2}(r-r_H)
\end{equation}
where
\begin{equation}
r_\pm=M\pm \sqrt{M^2-Q^2}, \;\;\;\;\; and\;\;\;\;\;r_H=r_+ ,
\end{equation}
where $M$ and $Q$ are the mass and the charge of the hole
respectively, and when $Q=0$, the metric of (\ref{metric}) reduces
into Schwarzschild's. Obvious, as $r\rightarrow \infty$,
$\sqrt{D(r)}\rightarrow 1-M/r$, and hence $E(r\rightarrow
\infty)_{QLE}=M$, i.e, the ADM-mass of the Reissner-Nordstr\"{o}m
black hole. And as $r=r_H$, we have $D(r_H)=0$ and
\begin{equation}\label{11}
E(r_H)_{QLE}=r_H/G
\end{equation}

We have mentioned in the Introduction that according to the
quantum principle, the quantum state energy $E$ and its conjugate
time $t$ can not be simultaneously measured exactly(Heisenberg
uncertainty principle). Namely, treating $E$ and $t$ as operators,
we have $[t,E]=i$. Given the relation eq.(\ref{11}), we must
conclude that the uncertainty of $E$ imply that of $r_H$, and then
we have
\begin{equation}\label{trn}
[t,r]|_{r=r_H}=il_p^2.
\end{equation}
Note this non-commutative relation  is independent on the
parameters of the black hole, and the corresponding uncertainty
relation  is $(\Delta t)(\Delta r)|_{r\sim r_H}\sim l_p^2$. In
other hands, due to quantum measurement effects, $r_H$ spread into
a range of $\{r_H-\Delta r,r_H+\Delta r\}$. Thus eq.(\ref{trn})
extends into
\begin{equation}\label{trnr}
[t,r]|_{r\in \{r_H-\Delta r,r_H+\Delta r\}}=il_p^2,
\end{equation}
where $\Delta r$ is a distance about Planck length scale, and will
be called as spread range of the horizon hereafter. With these
considerations, the original classical horizon became to the so
called {\it Quantum Horizon} which has a spread range and its
space-time is non-commutative in that range.

Non-commutative relation eq.(\ref{trnr}) means $(\Delta t)(\Delta
r) |_{r\sim r_H}\sim l_p^2$. This space-time uncertainty relation
can be re-explained  by other view.  In eq.(\ref{39}), QLE of the
black hole is the function of the radial coordinate $r$ and the
parameters of the black hole $r_H$. Now, we consider a quantum
fluctuation of the QLE
\begin{equation}
\Delta E_{QLE}={1\over G}\left(\Delta r-\sqrt{D(r)} \Delta
r-{r\over 2\sqrt{D(r)}}\left((r-r_H)\pa_r f(r) \Delta
r+f(r)(\Delta r-\Delta r_H)\right)\right)
\end{equation}
where $\Delta r$ and $\Delta r_H$ represent the fluctuations of
the radial and the ones of parameters respectively. Near the
horizon, i.e., $r= \lim_{\epsilon\rightarrow 0} (r_H+\epsilon)$,
we have
\begin{equation}\label{46}
\Delta E_{QLE}|_{r=(r_H+\epsilon)}={1\over G}\left(\Delta r
-r_H\sqrt{f(r_H)}(\Delta r-\Delta r_H)/\sqrt{\epsilon}\right).
\end{equation}
where $D(r)=f(r)(r-r_H)$ has been used. If requiring
 the fluctuation of the QLE  to be finite, then we have
\begin{equation}
 \Delta E_{QLE}|_{r\rightarrow r_H}={1\over G}\Delta r={1\over G}\Delta
 r_H.
\end{equation}
By using the Heisenberg uncertainty relation $(\Delta t)(\Delta
E_{QLE})\sim 1$, then we get $(\Delta t)(\Delta r)|_{r\sim
r_H}\sim G=l_p^2$ again.

\section{Quantum horizon and thermodynamics: the Schwarzschild Black hole}

In this section, we will construct a model based on the {\it
Quantum Horizon} to discuss the thermodynamic of the Schwarzschild
black hole.  The Schwarzschild metric is given by the
Eq.(\ref{Schwarzchild}). Treating the black hole as quantum state,
 the energy of quantum states is the QLE at the horizon, and then we
 have
the space-time non-commutative relation(\ref{trnr}). Now, we
construct a model with noncommutative $\phi$-field in the range of
$\{r_H -\Delta, r_H + \Delta\}$ and with commutative $\phi$-fields
in the ranges of $\{0, r_H -\Delta \}$ and  $\{r_H +\Delta,L\}$.
(for simplicity we use $\Delta$ to denote $\Delta r$ hereafter).
As the discussion in the Introduction, that $\Delta$ should be an
intrinsic quantity of the model, which characterizes the boundary
between the noncommutative space-time range and the commutative
space-time ranges, and it could be determined by the dynamics of
the model.  We rewrite eq.(\ref{trnr}) as follows
\begin{eqnarray}
&&[x^i,x^j]=i\Theta\varepsilon^{ij},\;\;\;
(i,j=0,1),\;\;\;(x^0=t,x^1=r),\;\;\;
 \Theta={l_p^2}, \\ \nonumber
&&[x^k,x^\mu]=0, \    \ (k=2,3;\mu=0,1,2,3)
\end{eqnarray}
where $\varepsilon^{ij}$ is antisymmetrical with
$\varepsilon^{01}=1$. The star product of two function $f(x)$ and
$g(x)$ is given by the Moyal formula (which has been studied in
\cite{Witten}\cite{Seiberg}):
\begin{equation}\label{sp1}
(f\star g)(x)= \exp\left[{i\over
2}\Theta\varepsilon^{ij}\frac{\partial}{\partial
x^i}\frac{\partial}{\partial y^j}\right]f(x)g(y)|_{y = x}.
\end{equation}
In terms of above notations, the action of the QFT-model in curved
space with {\it quantum horizon} is  as follows:
\begin{eqnarray}\label{Nac}
I&=&-{1\over 2}\int d^4 x \sqrt{-g} g^{\mu\nu} (\pa_{\mu}\phi
\star \pa_{\nu}\phi) \hskip0.4in (r_H-\Delta<r<r_H+\Delta),\\
\label{Nac1} I&=&-{1\over 2}\int d^4 x \sqrt{-g} g^{\mu\nu}
(\pa_{\mu}\phi \pa_{\nu}\phi) \hskip0.5in (r_H-\Delta<r,\;\;\;{\rm
or}\;\;r>r_H+\Delta),
\end{eqnarray}
where $\Delta$ is determined by the infinite red-shifting
condition we have discussed in above, i.e., by following equation
\begin{equation}\label{Delta}
\widetilde{g}_{tt}(r_H+\Delta)=0
\end{equation}
where $\widetilde{g}_{\mu\nu}$ is the effective metric describing
the effects of noncommutative equation of motion $\delta
I[\phi]=0$ in the range of $\{r_H-\Delta<r<r_H+\Delta \}$. We
address the QFT-model with the quantum horizon
(\ref{Nac})--(\ref{Delta}) is quite different from the QFT-model
with classical horizon ( for instance, the brick wall
model\cite{thooft85}). The fields in the action (\ref{Nac}) are
non-commutative. And this properties  are naturally  charactered
by the Planck length, i.e., $\Theta\equiv l_p^2$. Distinguishing
from other brick wall-type modes, there is no any free parameter
in the theory defined by eqs(\ref{Nac})-(\ref{Delta}).

We evaluate the start product in the action of (\ref{Nac}) using
eq.(\ref{sp1}) and cast the action in the ordinary product. By
this, the noncommutative effect can be absorbed into an equivalent
background metric. In other words, we first take the
semi-classical quantum effect into consideration. This effect is
then realized through the non-commutative geometry. Finally, this
effect is further through an effective background but in an
ordinary geometry. For a given energy mode, i.e., assuming the WKB
approximation wave function $\phi(t,r,\theta,\varphi)=e^{-it\omega
-i\int k(r) dr}Y_{lm}(\theta,\varphi)$, the effective metric can
be either read from the action (actually simpler) or from the
following equation of motion for the scalar field once the star
product is evaluated:
\begin{eqnarray}\label{NMT}
&&\pa_t(\sqrt {-g}g^{tt}\pa_t\phi)+\pa_r(\sqrt {-g}
g^{rr}\pa_r\phi)+\pa_\theta(\sqrt
{-g}g^{\theta\theta}\pa_\theta\phi)+\pa_\varphi(\sqrt
{-g}g^{\varphi\varphi}\pa_\varphi\phi)  \nonumber \\
&+&{1\over 2!}\left({i\Theta\over 2}\right)^2(\sqrt
{-g}g^{rr}),_{rr}\pa_t\pa_t\pa_r\pa_r\phi \\ \nonumber
&+&\sum_{n=1}^\infty{1\over 2n!}\left({i\Theta\over
2}\right)^{2n}(\sqrt {-g}g^{tt}),_{\underbrace{r\cdots
r}\limits_{2n}}\underbrace{{\pa_t\cdots \pa_t}}\limits_{2n+2}\phi
 =0
\end{eqnarray}
where $(\sqrt{ -g}g^{tt}),_{\underbrace{r\cdots r}\limits_{n}}$
stands for $\underbrace{{\pa_r\cdots \pa_r}}\limits_{n}(\sqrt{
-g}g^{tt})$, etc. For the scalar field with given energy $\omega$,
the above equation becomes
\begin{eqnarray}\label{NKG}
&&\left[-{\sin\theta r^3\over
r-r_H}-{\sin\theta\Theta^2\omega^2\over 4}-{\sin\theta r_H^3\over
r-r_H}\sum^\infty_{n=1}\left({\Theta \omega\over
2(r-r_H)}\right)^{2n}\right]\phi ,_{tt} + \nonumber \\
&&+ \left[ \sin\theta r(r-r_H)+{\sin\theta\Theta^2 \omega^2\over
4}\right] \phi ,_{rr}+(\sin\theta\phi ,_\theta),_\theta+{1\over
\sin\theta}\phi ,_{\varphi\varphi}=0
\end{eqnarray}
The noncommutative field action is nonlocal, which is include with
infinity orders of the differential operate. Under WKB
approximation, the $\Theta-$expansion terms in the equation of
motion (\ref{NKG}) reflect this non-local property. Note all
$\Theta-$dependent terms have been take into account. Using
formula
\begin{equation}\label{Nmo}
\sum^\infty_{n=1}\left({\Theta \omega\over
2(r-r_H)}\right)^{2n}={1\over 1-{\Theta^2\omega^2\over
4(r-r_H)^2}}-1,
\end{equation}
the compact expression of this summation of all  orders of
$\Theta$ is obtained. So, the nonperturbation effects of
$\Theta-$expansion are included.

  Comparing the noncommutative field equation of motion
Eq.(\ref{NKG}) to  the ordinary Klein-Gordon equation with curved
space-time metric $\tu{g}^{\mu\nu}$ (see Eq.(\ref{KG})), we obtain
 the effective metric $\tu{g}^{\mu\nu}$
\begin{eqnarray}
\sqrt{-\tu{g}}\tu{g}^{tt} &=& -{\sin\theta r^3\over r-r_H}\left[
1+{\Theta^2\omega^2(r-r_H)\over 4r^3}+{r_H^3\over r^3}\left(
{1\over 1-{\Theta^2\omega^2\over 4(r-r_H)^2}}-1\right)\right],  \\
\nonumber  \sqrt{-\tu{g}}\tu{g}^{rr} &=& \sin\theta \left[
r(r-r_H)+{\Theta^2\omega^2\over 4}\right], \\ \nonumber
\sqrt{-\tu{g}}\tu{g}^{\theta\theta} &=& \sin\theta, \\ \nonumber
\sqrt{-\tu{g}}\tu{g}^{\varphi\varphi} &=& {1\over \sin\theta}.
\end{eqnarray}
From the above, we can solve the effective metric as
\begin{eqnarray}\label{eff1}
\tu{g}_{\varphi\varphi}&=&\tu{g}_{\theta\theta}\sin^2\theta, \\
\label{eff2}
 \tu{g}_{tt}\tu{g}_{rr} &=& -1 ,\\ \label{eff3} \tu{g}_{tt} &=& -
\left( 1-{r_H\over r}\right) \sqrt{{1+{\tu{\Delta}^2(\omega)\over
r(r-r_H)}\over 1+{\tu{\Delta}^2(\omega)(r-r_H)\over
r^3}+{r_H^3\over r^3}\left[{1\over 1-{\tu{\Delta}^2(\omega)\over
(r-r_H)^2}}-1\right]}},\\ \label{eff4}  \tu{g}^2_{\theta\theta}
&=& \left[ r(r-r_H)+\tu{\Delta}^2(\omega)\right]\left[
r^2+rr_H+r_H^2+\tu{\Delta}^2(\omega)+{(r-r_H)r_H^3\over
(r-r_H)^2-\tu{\Delta}^2(\omega)}\right],
\end{eqnarray}
where we have set $\tu{\Delta}(\omega)=\Theta
\omega/2$.Substituting eq.(\ref{eff3}) into eq.(\ref{Delta}), we
obtain follows
\begin{equation}\label{Delta1}
\Delta(\omega)=\widetilde{\Delta}(\omega)\equiv {\Theta
\omega\over 2}.
\end{equation}
Thus the quantum horizon spread has been determined. We address
again that there are no additional parameters were introduced in
the above derivations, and it is essential that the quantum effect
causes the unexpected appearance of the new horizon at $r=r_H\pm
\Delta(\omega)$, at which $\widetilde{g}_{tt}$ vanishes. In
addition, note also that the $ \widetilde{g}_{\theta\theta} $
blows up at these two points which imply that the curvature scalar
vanishes there, too. Therefore they are regular. Note that, as
also discussed in \cite{thooft96}, the energy $\omega$ for the
scalar $\phi$ should not be too large, therefore
$\Delta(\omega)=\Theta \omega/2=l_p(l_p\omega/2)$ is not larger
than the Planck length $l_p$.

Using the WKB approximation, the square of $\phi$-field wave
number $k^2(r)$ in the spread range reads
\begin{eqnarray}
k^2(r_H-\Delta<&r<&r_H+\Delta)=  {\omega^2r^2\over
(r-r_H)^2+{(r-r_H)\Delta^2\over r}}  \\ \nonumber & \times&\left[
1+{(r-r_H)\Delta^2\over r^3}+{r_H^3\over r^3}\left({(r-r_H)^2\over
(r-r_H)^2-\Delta^2}-1\right)-{l(l+1)(r-r_H)\over
\omega^2r^3}\right]
\end{eqnarray}
The square of $\phi$-wave number $k^2(r)$ besides the spread range
can be read from the Eq.(\ref{kk}). By using $S$-wave approximate,
the wave number $k$ reduces to be
\begin{eqnarray}\label{sNk}
k(r)&=&\pm i{\omega
r_H\over\Delta\sqrt{1-x^2}}\sqrt{{x+3\xi(x^2-1)+3\xi^2x(x^2-1)+\xi^3(x^4-1)\over
x(1+x\xi)(1+{\xi\over \xi x^2+x})}},\\ \nonumber
&&{\rm in\; the\; range}:(r_H-\Delta<r<r_H+\Delta);  \\
k(r)&=& \pm {\omega r\over r-r_H} \\ \nonumber && {\rm in\;the\;
range}:(r<r_H-\Delta \;\;\; {\rm or} \;\;\; r>r_H+\Delta).
\end{eqnarray}
where $x=(r-r_H)/\Delta$ and $\xi=\Delta /r_H$. Note, the mass of
macro black hole we concern is much larger than Planck mass, hence
$\xi\ll 1$. Now, ingoing and outgoing wave function of the $\phi$
with energy $\omega$ are follows
\begin{eqnarray}
\phi^{in}_\omega&=& \exp\left[ -i\omega\left(
t+\int^r_{r_H+\Delta}{rdr\over r-r_H}\right)\right],\\ \nonumber
\phi^{out}_\omega(r>r_H+\Delta) &=& A_\omega\exp\left[ -i\omega
t+i\int^r_{r_H+\Delta} {\omega rdr\over r-r_H}\right]\\ &=&
A_\omega\phi^{in}_\omega\exp\left[ 2i\int^r_{r_H+\Delta} {\omega
rdr\over r-r_H}\right], \\ \phi^{out}_\omega(r<r_H-\Delta)&=&
A_\omega\phi^{in}_\omega\exp\left[ 2i
\int^{r_H-\Delta}_{r_H+\Delta} k(r)dr\right]\exp\left[
2i\int^r_{r_H-\Delta} {\omega rdr\over r-r_H}\right].
\end{eqnarray}
Using Eq.(\ref{sNk}), the first integration in r.h.s. of equation
can be calculated,
\begin{equation}
2i\int^{r_H-\Delta}_{r_H+\Delta}k(r)dr=\pm \omega
r_H\int^1_{-1}{dx\over
\sqrt{1-x^2}}\sqrt{{x+3\xi(x^2-1)+3\xi^2x(x^2-1)+\xi^3(x^4-1)\over
x(1+x\xi)(1+{\xi\over \xi x^2+x})}}
\end{equation}
Since $\xi\ll 1$, we have
\begin{equation}\label{xi}
2i\int^{r_H-\Delta}_{r_H+\Delta}k(r)dr\simeq \pm(2\pi r_H\omega\pm
i2\pi r_H\omega\xi +3\pi r_H\omega\xi^2+\mathcal{O}(\xi^3))\simeq
\pm 2\pi r_H\omega(1\pm i\xi)
\end{equation}
where the terms of $\mathcal{O}(\xi^2)$ have been neglected  due
to $\xi\ll 1$, whose effects will be briefly discussed in the end
of this section. Taking the positive sign, we obtain the absolute
value of ratio of the outgoing wave function's amplitude outside
the horizon to the one inside as follows
\begin{equation}
\left|{\phi^{out}_\omega(r>r_H+\Delta)\over
\phi^{out}_\omega(r<r_H-\Delta)}\right| = e^{-2\pi r_H \omega}.
\end{equation}
According to the Eqs.(\ref{ratio})-(\ref{ratio2}), we have
\begin{equation}
T_{BH}={1\over 4\pi r_H},
\end{equation}
then, we get the Hawking temperature of the Schwarzschild black
hole.

   From the effective metric $\tu{g}_{\mu\nu}$ of the
Eqs.(\ref{eff1})-(\ref{eff4}), it is now not difficult to
understand the boundary condition (\ref{bc}) we given earlier.
Because of the appearance of the new horizon, an observer outside
the horizon can not detect the $\phi_\omega$ at
$r<r_H+\Delta(\omega)$. So, the number of states below energy
$\omega$ is
\begin{equation}\label{gg}
g(\omega)=N\int dl(2l+1)\int^L_{r_H+\Delta(\omega)}
dr\sqrt{k^2(r,l,\omega)}
\end{equation}
where $k^2(r,l,\omega)$ is given by the Eq.(\ref{kk}). Comparing
Eq.(\ref{gg}) with the Eq.(\ref{g}), the $g(\omega)$ in our model
is quite different from the BWM: The lower bound of the integral
in eq.(\ref{gg}) is $r_H+\Delta(\omega)$, but in BWM, this lower
bound  is $r_H+h$, which is independent on the energy of the
$\phi$, and $h$ is put in by the hand instead of by the dynamics.
The free energy then reads
\begin{eqnarray}\nonumber
\pi\beta F &=& \int dg(\omega) \ln\left( 1-e^{-\beta\omega}\right)
\\ \nonumber &=& -\int^\infty_0 d\omega {\beta g(\omega)\over
e^{\beta\omega}-1}\\ &=& -{2N\over 3}\int^\infty_0 d\omega
{\beta\omega^3\over
e^{\beta\omega}-1}\int^L_{r_H+\Delta(\omega)}dr{r^4\over
(r-r_H)^2}.
\end{eqnarray}
The dominant contribution from the event horizon to $F$ is
\begin{equation}
F\approx -{8N\zeta(3)r_H^4\over 3\pi \beta^3l_p^2},
\end{equation}
where we have used $\Delta(\omega)=\omega l_p^2/2$ and $\zeta(3)$
is Riemann $\zeta$-function,
$\zeta(3)=\sum^\infty_{n=1}1/n^3\approx 1.202$.The contribution of
the horizon to the entropy $S$ are
\begin{eqnarray}
 S=\beta^2{\pa F\over \pa \beta} ={8N\zeta(3)r_H^4\over
\pi\beta^2l_p^2}.
\end{eqnarray}
If we set $\beta =4\pi r_H$ in above equations, we have
\begin{eqnarray}\label{U}
 \label{S} S = {N\zeta(3)A\over
8\pi^4l_p^2}
\end{eqnarray}
where $A=4\pi r_H^2$ is the area of the black hole. It is
remarkable that the entropy of (\ref{U}) is proportional to $A$.
This indicates we are indeed on the right track.

On the other hand, the thermodynamics require the black hole
entropy has to be normalized to the Bekenstein-Hawking expression,
i.e., $S=S_{BH}=A/(4l_p^2)$. Then, we obtain the number $N$ as
follows
\begin{equation}\label{NN}
N=2\pi^4/\zeta(3)\approx 162.
\end{equation}
 In the next section, we will show that the result of (\ref{NN})
  is not only valid for the Schwarzschild black hole, it is also right for all static
 spherical black holes.

To reveal the physical meaning of the brick wall thickness $h$
introduced by 't Hooft, we calculate
 the statistical average value of  $\Delta(\omega)$
\begin{equation}\label{ad}
\overline{\Delta(\omega)}={l_p^2\over
2}\overline{\omega}={l_p^2\over 2}{\int^\infty_0{\omega
dg(\omega)\over e^{\beta\omega}-1}\over\int^\infty_0
{dg(\omega)\over e^{\beta\omega}-1}}={3\zeta(3)l_p^2\over\pi^3
r_H}.
\end{equation}
Comparing eq.(\ref{ad}) with (\ref{h}), we find
\begin{equation}
h=\alpha \overline{\Delta(\omega)},
\end{equation}
where $\alpha= N'\pi^2/(1080\zeta(3))$ is a constant. Therefore we
conclude
 the "brick wall" thickness $h$ represents the
statistical average effects of the quantum horizon spread range
$\Delta(\omega)$.

  Finally in this section, we argue that the effects of
$\mathcal{O}(\xi^2)$ in eq.(\ref{xi}) raise the effective
temperature of the hole as it radiates. Namely, the
$\xi$-dependency in the eq.(\ref{xi}) should be thought as its
thermal statistical average $\overline{\xi}$-dependency. like
eq.(\ref{ad}),
$\overline{\xi^2}=l_p^4\overline{\omega^2}/(4r_H^2)=1/160\times(l_p/r_H)^4$,
then the effective temperature for the hole is
\begin{equation}
T_{eff}=T_{BH}\left(1-{3l_p^4\over 320r_H^4}\right)
\end{equation}
where the $T_{BH}$ is Hawking temperature and the second term in
the right-hand-side represents a correction to the temperature due
to the space-time non-commutative property near the event horizon.
Obviously, this correction to the $T_{BH}$ is tiny as $r_H\gg
l_p$, and hence it can be ignored indeed. The corrections of
$\mathcal{O}(\xi^N)$ with $N>2$ can be analyzed likewise and they
are also ignorable as $r_H\gg l_p$.

\section{Quantum horizon and thermodynamics: general static spherical black hole}

In this section, we study our QFT model with quantum horizon for
general static black holes. The static spherical black hole's
metric generically can be written as the eq.(\ref{metric}), and
the Schwarzschild black hole is a special case of this metric. The
calculation process is similar to the previous section, except the
metric is different. As  discussed in above , the noncommutative
range near the horizon is not bigger than the Planck length.
Hence, to the metric near the horizon the function $D(r)$ in
(\ref{metric}) can be approximately written  as follows
\begin{equation}\label{D1}
D(r)=f(r_H)(r-r_H)+\mathcal{O}((r-r_H)^2)
\end{equation}
This approximation corresponds to $\xi\rightarrow 0$ in the
previous section calculations for  the Schwarzschild black hole.
Substituting the metric to the eq.(\ref{NMT}), and taking
WKB-approximation, the equation motion of the scalar field with
energy $\omega$ is
\begin{eqnarray}\label{em}
&-&{\sin\theta r^2 \over f(r_H)(r-r_H)}\left(
1-{\widetilde{\Delta}(\omega)^2\over (r-r_H)^2} \right)^{-1}
\phi,_{tt}\\ \nonumber &&+ \sin\theta r^2
f(r_H)(r-r_H)\left(1+{2\widetilde{\Delta} (\omega)^2 \over
r_H(r-r_H)}\right)\phi ,_{rr}\\ \nonumber &&+\pa_\theta
(\sin\theta \pa_\theta \phi)+{1\over \sin\theta} \phi ,_{\varphi
\varphi}=0
\end{eqnarray}
where $ \widetilde{\Delta}(\omega)\equiv\Theta \omega /2$, and the
following formula has been used
\begin{equation}\label{Nmo2}
\sum_{n=0}^\infty \left( {\widetilde{\Delta}( \omega)\over
r-r_H}\right)^{2n}=\left( 1-{\widetilde{\Delta}(\omega)^2\over
(r-r_H)^2} \right)^{-1}.
\end{equation}
Comparing eq(\ref{em}) with the standard Klein-Gordon equation in
the curved space with effective metric $\widetilde{g}_{\mu\nu}$,
we have\cite{by03}
\begin{eqnarray}\label{eff}
\sin ^2 \theta \widetilde{g}_{\theta \theta}&
=&\widetilde{g}_{\varphi\varphi},\\ \label{eff2}
\widetilde{g}_{tt}\widetilde{g}_{rr} &=& -1 \\ \label{eff3}
\widetilde{g}_{tt}&=&-f(r_H)(r-r_H)\sqrt{\left(1-{\widetilde{\Delta}
^2 \over
(r-r_H)^2}\right)\left(1+{2\widetilde{\Delta}^2\over r_H (r-r_H)}\right)} \\
\label{eff4} \widetilde{g}_{\theta \theta} &=&
r^2\sqrt{1+{2\widetilde{\Delta}^2\over r_H(r-r_H)}\over
1-{\widetilde{\Delta}^2\over (r-r_H)^2}}
\end{eqnarray}
Substituting eq.(\ref{eff3}) into eq.(\ref{Delta}), we obtain
\begin{equation}\label{Delta1}
\Delta(\omega)=\widetilde{\Delta}(\omega)\equiv {\Theta
\omega\over 2}.
\end{equation}

Using WKB-wave function $\phi=\exp (-i\omega t -i\int k(r)dr)$,
$S-$wave approximation, the equation of motion (\ref{em}) and the
usual dispersion relation due to the commutative action
(\ref{Nac1}) , the wave number $k(r)$ with given energy $\omega$
in the both noncommutative and commutative ranges reads
respectively
\begin{eqnarray}\label{cwn}
k(r)&=&\pm i\omega f^{-1}(r_H)(r-r_H)^{-1} \sqrt{{\Delta(\omega)
\over
(r-r_H)^2}-1},\\ \nonumber
&&{\rm in \; the \;range:\;\;}(r_H-\Delta (\omega)<r<r_H +\Delta(\omega))\\
\label{cwn1}
 k(r)&=&\omega f(r)^{-1}(r-r_H)^{-1}\\ \nonumber
 && {\rm in\;the\;range:} (0<r< r_H-\Delta (\omega))\;\; {\rm or}
 \;\; (r>r_H
+\Delta(\omega)).
\end{eqnarray}
Note the wave number $k$ in the  noncommutative range is
imaginary. This means that the quantum tunnelling occurred in this
noncommutative range. The ingoing and outgoing wave functions are
as follows
\begin{eqnarray}
\phi^{in}_\omega&=& \exp\left[ -i\omega\left(
t+\int^r_{r_H+\Delta}{dr\over f(r)(r-r_H)}\right)\right],\\
\nonumber \phi^{out}_\omega(r>r_H+\Delta) &=& A_\omega\exp\left[
-i\omega t+i\int^r_{r_H+\Delta} {\omega dr\over f(r)(r-r_H)}\right]\\
&=& A_\omega\phi^{in}_\omega\exp\left[ 2i\int^r_{r_H+\Delta}
{\omega dr\over f(r)(r-r_H)}\right], \\
\phi^{out}_\omega(r<r_H-\Delta)&=&
A_\omega\phi^{in}_\omega\exp\left[ 2i
\int^{r_H-\Delta}_{r_H+\Delta} k(r)dr\right]\exp\left[
2i\int^r_{r_H-\Delta} {\omega dr\over f(r)(r-r_H)}\right].
\end{eqnarray}

The ratio between the amplitude of outgoing wave function of
$\phi$ outside the hole and one inside it is
$\exp({2i\int^{r_H-\Delta}_{r_H+\Delta} k(r)dr})$, which describes
the the quantum tunnelling effects of $\phi$ passing the
non-commutative range of the black hole. Using eq.(\ref{cwn}), the
following integration can be be calculated
\begin{eqnarray}\label{nkk}
Im\left( 2\int k(r)dr\right)&=&
2i\int^{r_H-\Delta(\omega)}_{r_H+\Delta(\omega)}k(r)dr=\pm
{2\omega \over f(r_H)}\int ^{+1}_{-1} {dx\over \sqrt{1-x^2}}\\
\nonumber & =&\pm {2E_\phi\over f(r_H)} \arcsin (x)|^1_{-1}=\pm
{2\pi \omega \over f(r_H)}
\end{eqnarray}
where $x=(r-r_H)/\Delta(\omega)$ is dimensionless parameter, and
the result of (\ref{nkk}) is independent of the non-commutative
parameter $\Theta$. By means of the theory of
Dammur-Ruffini\cite{Ruffini} and Sannan\cite{Sannan}, the black
hole temperature $T_{BH}$ can be derived from the ratio between
the amplitude of outgoing wave function outside the hole and one
inside the hole, i.e., $\exp(|{2i\int^{r_H-\Delta}_{r_H+\Delta}
k(r)dr}|)=\exp (E_\phi /(2T_{BH}))$. Thus, using (\ref{nkk}), we
get the desired result as follows
\begin{equation}\label{T1}
T_{BH}={f(r_H) \over 4\pi}={\kappa\over 2\pi},
\end{equation}
where $\kappa=f(r_H)/2$ is the surface gravity on event horizon of
the given metric (\ref{metric}). Then, the standard Hawking
temperature has been reproduced successfully.

  Because of the appearance of the new horizon, an observer cannot
detect the $\phi$ at $r\leq r_H +\Delta(\omega)$. Just we have
discussed earlier that the field has an infinite ret-shift at
$r=r_H +\Delta(\omega)$. We can now follow the method of the above
section to evaluate the black hole entropy. The WKB wave number
$k(r)$ read from the action eq.(\ref{Nac1}) is
\begin{equation}
k^2(L>r>r_H+\Delta(\omega))=f^{-1}(r)(r-r_H)^{-1}\left(
f^{-1}(r)(r-r_H)^{-1} \omega^2 -l(l+1)r^{-2} \right)
\end{equation}
The number of states below energy $E_\phi$ is
\begin{equation}
g(\omega)=N \int dl(2l+1)\int^L_{r_H+\Delta(\omega)} dr
\sqrt{k^2(l,r,\omega)}
%= {2N\over 3f^2(r_H)}\int
%^L_{r_H+\Delta(E_\phi)} dr {E_\phi^3 r^2\over (r-r_H)^2}
\end{equation}
The free energy then reads
\begin{equation}
\pi \beta F= \int dg(\omega) \ln \left(1-e^{-\beta \omega}
\right)= -\int^\infty_0 d \omega {\beta g(\omega)\over e^{\beta
\omega}-1}
\end{equation}
where $\beta =1/T_{BH}$ is inverse temperature. The dominant
contribution from the even horizon to $F$ is
\begin{equation}
F\approx - {8N\zeta (3) r_H^2\over 3\pi f^2(r_H)l_p^2 \beta^3}
\end{equation}
The contribution of the horizon to  the entropy $S$ are
\begin{eqnarray}
 \label{entropy} S=\beta^2{\pa F\over \pa \beta}={N\zeta(3)A\over 8\pi^4 l^2_p}.
\end{eqnarray}
where $A=4\pi r_H^2$ is the horizon area. So we reproduce the
correct relation of $S\propto A$.normalized to the
Bekenstein-Hawking expression, i.e., $S=S_{BH}=A/(4l_p^2)$. Then,
from the eq.(\ref{entropy}) we obtain the field number $N$  as
follows
\begin{equation}\label{N}
N=2\pi^4/ \zeta(3)\approx 162.
\end{equation}
The number $N$ is  independent of the parameters of the black
hole, and same as eq.(\ref{NN}).

\section{summery and discussion}
In summery: A semi-classical reasoning leads to the
non-commutativity of space and time coordinates near the horizon
of static non-extreme black hole, and renders the classical
horizon spreading to {\it Quantum Horizon} . In terms of the
background metric of the black hole with the {\it Quantum Horizon}
, a quantum field theory in curved space without ultraviolet
divergency near the horizon is formulated. In this formulism, the
black hole thermodynamics is reproduced correctly without both
ambiguity and additional hypothesis in the deriving the hole's
Hawking radiations and entropies, and a new interesting prediction
on the number of radiative field modes  $N$  is provided.
Specifically, the main results are follows: 1, Hawking radiations
rightly emerge as an effect of quantum tunneling through the
quantum horizon, and hence the ambiguities due to going across the
singularity on the classical horizon were got rid of; 2, 't
Hooft's brick wall thickness hypothesis and the boundary condition
imposed for the field considered in his brick wall model were got
rid of also, and related physics has been interpreted; 3, The
present theory is parameter free. So, the theory has power to
predict the multiplicity $N$ of radiative field modes according to
the requirement of normalization of Hawking-Bekenstein entropy. It
has been found that $N\simeq 162$.

Finally, two discussions are in order:

 1)  We conclude in this paper
that the black hole entropy and the Hawking temperature have been
produced successfully and simultaneously in the QFT-model in
curved space with quantum horizon. Physically, this implies that
this model not only  can be used to rightly count the micro-states
of the black hole (corresponding to derive its entropy), but also
can  be used to rightly calculate the quantum tunnelling
amplitudes across the horizon for the particles creating by strong
gravitational fields (corresponding to derive the Hawking
temperature). Thus we have achieved to bridge those two black hole
physics phenomena consistently by one semi-classical quantum field
theory. All puzzles caused by the singularity associating with the
classical horizon mentioned in Section II have been solved by
considering the quantum effects on the horizon. In the theory,
there are no additional cutoff and any adjustable parameters that
are introduced. Thus, the successes of our theory directly support
the quantum horizon conception suggested by us in this paper.

 2) As we have mentioned in the above that the
thermodynamics requires the black hole entropy has to be
normalized to the Bekenstein-Hawking expression, i.e.,
$S=S_{BH}=A/(4l_p^2)$. Then, from the eq.(\ref{entropy}) we obtain
the field number $N$  as follows
\begin{equation}\label{NNN}
N=2\pi^4/ \zeta(3)\approx 162,
\end{equation}
where $N$ fails to be a integer both because some approximations
(e.g., WKB-approximation etc) in the state-counting are used in
the above and because some fine adjustments to the entropy due to
real world particle's charge and spin etc are ignored in this
paper. Such sort of errors
 should be rather small. Eq.(\ref{NNN}) is a good
 estimate to the field number of $\phi-$multiplet
 in the model.

It is well known that  the particle-antiparticle(with negative
energy)-pair creation effects in physical vacuum induced by the
strong classical gravitational fields will occur near the horizon
and will lead to the Hawking radiations. In absence of a full
quantum gravity theory, we could regard the vacuum as a
particle-antiparticle sea. Because  the masses of all particles in
real world physics should be much less the Planck mass  and
because also the gravitational interactions near the black hole
horizon are so  strong  that the interactions in the ordinary
particle dynamics can be ignored, the matter quantum dynamics near
the black hole will reduce  to action of free massless scalar
fields within the background gravitational fields like
eqs(\ref{Nac})(\ref{Nac1}). In this  picture, the charges, spins
and all intrinsic character of the particles and the interactions
between them are all disappeared (or ignorable), but the field
modes should keep to be unchangeable, and then the number of
degree of freedom of the fields are same as one in the case
without such kind of extremely strong background gravitational
fields, i.e., the original dynamical Lagrangian case.
Consequently, through counting the black hole's entropy, we could
get the field mode number $N$ in the QFT model with quantum
horizon, and then the number of field degree of freedom in the
underlying dynamics will be determined. Here the interesting
criterion is that those two numbers must be equal each other in
principle, because all particle-pair created from the vacuum must
belong to the underlying matter theory and vice-versa, i.e., all
particles in the theory must be created by the extremely strong
gravitational interactions. Therefore, the mode number $N$ should
be independent of the parameters of black hole, i.e., $N$ displays
an intrinsic character of vacuum. That the black hole physics
result could be helpful for searching the new physics beyond the
standard model is remarkable.

Back to our result eq.(\ref{NNN}). Surely, $N\simeq 162$ is much
larger than the number of degree of freedom of the Standard Model
(SM) , which is 82 (see the Table I). Therefore, at Planck scale
(or near Planck scale), SM fails, and must be ruled out. When one
counted the number of degree of freedom of the Minimal
Super-symmetric Standard Model (MSSM), it has been found that this
dynamical degree number is 164 (see Table I) , which is very close
to our prediction $N=162$. Hence, our model favor to support MSSM
as a new physics theory beyond SM.

\begin{center}
\begin{table}
%\caption{Multiple number of fields in minimal supersymmetric
%standard model \cite{Haber} }
\begin{tabular}{|l|l|c|l|l|c|}\hline
   \multicolumn{6}{|c|}{\bfseries Table I: Multiple number of fields in }\\
   \multicolumn{6}{|c|}{\bfseries minimal supersymmetric standard model\cite{Haber}} \\
\hline
Normal$\;\;\;\;\;\;\;\;$ & Name & N& Suppersymme- & Name & N \\
particles &    &   &tric partners & &  \\
\hline \hline $q=u,d,s,$ & quark & & $\tu{q}_L$, $\tu{q}_R$ &
scalar-quark&
\\ $\;\;\;\;\;\;\; c,b,t$ &$\;$($\times 3$ color) & 36 & &$\;\;\;\;\;\;$ ($\times 3$ color)&36 \\
\hline $l=e,\mu,\tau$ & lepton & 6 & $\tu{l}_L$, $\tu{l}_R$ &
scalar-lepton & 6 \\ \hline
 $\nu = \nu_e , \nu_\mu , \nu_\tau $ &  neutrino & 6 & $\tu{\nu}$
 & scalar-neutrino & 6 \\ \hline
 g & gluon & 16 & $\tu{\rm g}$ & gluino & 16 \\ \hline
 $W^{\pm}$ &  & 4 & $\tu{W}^\pm $ & wino & 4 \\ \hline
 $Z^0$ & & 2& $\tu{Z}^0$ & zino & 2 \\ \hline
 $\gamma$ & photon & 2 & $\tu{\gamma}$ & photino & 2 \\ \hline
 $H^+_1 \;\;\;\;\;\; H^0_1$ & & &$\tu{H}^+_1 \;\;\;\;\;\;
 \tu{H}^0_1$& & \\
$H^+_1 \;\;\;\;\;\; H^0_1$ & higgs & 8 &$\tu{H}^+_1 \;\;\;\;\;\;
\tu{H}^0_1$ & higgsino & 8 \\ \hline $g_{\mu\nu}$ & graviton & 2 &
$\tu{g}_{\mu\nu}$ & gravitino & 2 \\ \hline
\end{tabular}
\end{table}
\end{center}

\begin{center}
{\bf ACKNOWLEDGMENTS}
\end{center}
This work is partially supported by NSF of China 90103002 and the
PhD Program Fund of Chinese Education Ministry. The authors wish
to thank Dao-Neng Gao, Jiliang Jing,
 Jian-Xin Lu, and Shuang-Qing Wu  for their
stimulating discussions.


\begin{thebibliography}{99}
\bibitem{by2} In the ref.\cite{by03}, we set $E$ to be ADM-mass.It is not approperiate.
\end{thebibliography}
\end{document}